\documentclass[a4paper,12pt]{article}
\usepackage{graphicx}

\newcommand{\beq}{\begin{equation}}
\newcommand{\eeq}{\end{equation}}
\def\bea{\begin{eqnarray}}
\def\eea{\end{eqnarray}}

\def\journal#1#2#3#4{{\it #1} {\bf #2} (#3) #4}
\def\epj{Euro. Phys. Jour.}
\def\prl{Phys. Rev. Lett.}
\def\pl{Phys. Lett.}
\def\np{Nucl. Phys.}
\def\ptp{Prog. Theor. Phys.}

\def\pr{Phys. Rev.}

\def\ijmp{Int. Jour. Mod. Phys.}

\def\l{\ell}

\def\b{{\cal B}}
\def\f{{\cal C}}

\def\h0t{{\tilde{h}^0}}

\def\bqll{B_q \rightarrow \l^+ \, \l^-}

\def\bqtt{B_q \rightarrow \tau^+ \, \tau^-}
\def\bdll{B_d \rightarrow \l^+ \, \l^-}
\def\bsll{B_s \rightarrow \l^+ \, \l^-}
\def\bstt{B_s \rightarrow \tau^+ \, \tau^-}

\def\bsmm{B_s \rightarrow \mu^+ \, \mu^-}
\def\bdmm{B_d \rightarrow \mu^+ \, \mu^-}

\def\b{{\cal B}}

\def\alp{{\cal A}_{\rm LP}}

\def\h{{\cal H}}

\def\faa{\f_{\rm AA}}
\def\fpp{\f_{\rm PP}}
\def\fps{\f_{\rm PS}}

\def\ml{m_\l}

\title{ Longitudinal Polarization Asymmetry of Leptons \\
in pure Leptonic $B$ Decays}
\author{
        {\bf L. T. Handoko}$^{1,2}$\thanks{
        E-mail : handoko@lipi.fisika.net, handoko@fisika.ui.ac.id},  \hspace{2mm}        
        {\bf C. S. Kim}$^3$\thanks{
        E-mail : cskim@mail.yonsei.ac.kr,~~http://phya.yonsei.ac.kr/\~{}cskim/}
        \hspace{2mm} and \hspace{2mm}
      {\bf T. Yoshikawa}$^4$\thanks{
        E-mail : tadashi@physics.unc.edu} \\
        \vspace*{1mm} \\
        $^1$Pusat Penelitian Fisika, LIPI\thanks{http://lipi.fisika.net} \\
        Kompleks PUSPIPTEK Serpong, Tangerang 15310, Indonesia\\
        \vspace*{0.3mm} \\
        $^2$Jurusan Fisika FMIPA, Universitas Indonesia \\
        Depok 16424, Indonesia\\
        \vspace*{0.3mm} \\
        $^3$Department of Physics and IPAP, Yonsei University \\
        Seoul 120-749, Korea \\
        \vspace*{0.3mm} \\
        $^4$Department of Physics and Astronomy, 
        University of North Carolina \\
        Chapel Hill, NC 27599-3255, USA
        }
\date{\today}

\begin{document}

\maketitle
\begin{picture}(0,0)
       \put(310,420){FISIKALIPI-01005}
       \put(310,405){FIS-UI-TH-01-01}
       \put(310,390){IFP-803-UNC}
\end{picture}

\thispagestyle{empty}

\begin{abstract}

\noindent 
Longitudinal polarization asymmetry of leptons in  $\bqll$
($q = d, s$ and $\l = e, \mu, \tau$)
decays is investigated.
The analysis is done in a general manner by using the effective
operators approach. It is shown that the longitudinal polarization
asymmetry would provide a direct search for the  scalar and pseudoscalar 
type interactions, which are induced in all variants of Higgs-doublet models.

\end{abstract}

\clearpage

It has been already pointed out by several authors 
\cite{logan,logan2,huang2hd,kruger} that
the pure leptonic $B$ decays $\bqll$ ($q = d, s$ and $\l = e, \mu, \tau$)
are very good probes to test new physics beyond the standard
model (SM), mainly to reveal the Higgs sector. 
Those previous works were focused on the contributions
induced by the scalar and pseudoscalar interactions realized in 
Higgs-doublets models. Within the SM, the decays are
dominated by the $Z-$penguin and the box diagrams, which are
helicity suppressed. We note that Higgs-doublet models
can generally enhance the branching ratio significantly.
Also, as discussed in recent works, the decays are strongly
correlated with the semi-leptonic $B$ decays \cite{kruger} and
even with the muon anomalous magnetic moment \cite{nierste}.
Experimentally, it is expected that present and the forthcoming
experiments on the $B-$physics ($B-$factories) can probe the
flavor sector with high precision \cite{bfactory}.

If we detect large discrepancy between the theoretical estimation of
the decay branching fractions and the actually observed experimental
results, then  this could be either an evidence of new physics
or of our lack of knowledge of the decay constants of $B$ mesons, $f_{B_q}$.
Therefore, the main interest would be a direct observation of new physics contributions
belonging to the non-SM interactions, $i.e.$ the scalar and pseudoscalar
interactions, because within the SM the decay is only through the axial vector
interactions.
In this letter, we propose a new
observable, namely the longitudinal polarization asymmetry 
of leptons ($\alp$) in $\bqll$
($q = d, s$ and $\l = e, \mu, \tau$) decays.
Though the measurement may be very difficult and challenging, 
we point out that this observable is very sensitive to those  
non-SM new interactions, and provides a direct evidence of their existence. 
We notice that the idea of measuring $\alp$ and CP--violation in 
$K_L \rightarrow \mu^+ \, \mu^-$ decay to 
look for new physics has been previously considered 
in several papers \cite{kll}. 
However, we would like to mention that those observables are quite different 
in the $B$ decay system \cite{bll,huangCP}: 
In the $K$ system the initial CP--eigenstate 
can be determined 
due to large lifetime difference of $K_{L,S}$, while 
such determination is not possible in the case of $B$ meson system. 
Therefore, we cannot 
discuss the $\bqll$ decays in the same manner as those previous references.  

Taking into account all possible 4-fermi operators which
could contribute to  $\bqll$, these 
processes are governed by the following effective Hamiltonian
\cite{fkmy},
\begin{eqnarray}
  \h_{\rm eff} & = & - \frac{G_F \alpha}{2 \sqrt{2} \pi}
  \left( V_{tq}^\ast V_{tb} \right) \,
  \left\{ 
  \faa 
  (\bar{q} \, \gamma_\mu \gamma_5 \, b)
  (\bar{\l} \, \gamma^\mu \gamma_5 \, \l)
  \right. \nonumber \\
  & & \left. \; \; \; \; \; \; \; \; \; \; \; \; \; \; 
  +
  \fps 
  (\bar{q} \, \gamma_5 \, b)
  (\bar{\l} \, \l)
  +
  \fpp 
  (\bar{q} \, \gamma_5 \, b)
  (\bar{\l} \, \gamma_5 \, \l)
  \right\} \; , 
  \label{eqn:heff1}
\end{eqnarray}
by normalizing all terms with the overall factors of the SM.
In particular, within the SM one has
$\fps^{\rm SM} = \fpp^{\rm SM} \simeq 0$ and  
$\faa^{\rm SM } = {Y(x_{t_W})}/{\sin^2 \theta_W}$,
where $Y(x_{t_W})$ is the Inami-Lim function \cite{inamilim}
with $x_{t_W} = ({m_t}/{M_W})^2$. The contributions
proportional to $m_{d,s}$ are neglected, and  the neutral Higgs
contributions in $\fps^{\rm SM}$ and $\fpp^{\rm SM}$ are 
proportional to ${(m_\l m_b)}/{m_W}^2$, and therefore also neglected.

After using the PCAC ansatz to derive the relation between the operators,
the most general matrix element for the decay is
\begin{eqnarray}
{\cal M} & = & i f_{B_q}\frac{G_F \alpha }{2 \sqrt{2} \pi }V_{tq}^*V_{tb}
             \left[ \left( 2 m_\l \faa - \frac{m_{B_q}^2}{m_b + m_q} \fpp 
                    \right) \bar{\l} \, \gamma_5 \, \l 
                   -   \frac{m_{B_q}^2}{m_b + m_q} \fps \bar{\l} \, \l 
             \right].
\label{eqn:amp}
\end{eqnarray}
Using Eq. (\ref{eqn:amp}),
the branching ratio for $\bqll$ becomes
\begin{eqnarray}
  \b(\bqll)  & = & \frac{G_F^2 \, \alpha^2}{64 \pi^3} \,
  \left| V_{tq}^\ast V_{tb} \right|^2 \, \tau_{B_q} f_{B_q}^2 \,
  m_{B_q} \, \sqrt{1 - \frac{4 \ml^2}{m_{B_q}^2}} 
   \nonumber \\
  & &   \times   \left[
  \left| 2 m_\l ~\faa  - \frac{m_{B_q}^2}{m_b + m_q} \fpp\right|^2
  + \left( 1 - \frac{4 m_\l^2}{m_{B_q}^2}\right)
  \left| \frac{m_{B_q}^2}{m_b + m_q} \fps\right|^2
   \right],
  \label{eqn:br}
\end{eqnarray}
where $\tau_{B_q}$ is the life-time of $B_q$ meson. The QCD
correction in this decay mode is remarkably negligible.
As can be easily seen,
the significant branching ratio within the SM could
be expected only for $\l = \tau, \mu$ due to the lepton mass dependence.

We now define an observable using
the lepton polarization.  Since in the dilepton rest frame
we can  define only one direction,
the lepton polarization vectors in each lepton's rest frame
are defined as
\begin{equation}
  \bar{s}^\mu_{\l^\pm} = \left( 0, \pm \frac{\mathbf{\rm p_-}}{|\mathbf{\rm p_-}|} \right) \; ,
\end{equation}
and  in the dilepton rest frame they
are boosted to
\begin{equation}
  s^\mu_{\l^\pm}  = \left( \frac{|\mathbf{\rm p_-}|}{m_\l},
    \pm \frac{E_\l \mathbf{\rm p_-}}{m_\l |\mathbf{\rm p_-}|} \right) \; ,
\end{equation}
where $E_\l$ is the lepton energy.
Finally the longitudinal polarization asymmetry of the final leptons in $\bqll$
is defined as follows;
\begin{equation}
  \alp^\pm \equiv 
  \frac{
    \left[ \Gamma(s_{\l^-},s_{\l^+}) + \Gamma(\mp s_{\l^-},\pm s_{\l^+}) \right] - 
    \left[ \Gamma(\pm s_{\l^-},\mp s_{\l^+}) + \Gamma(-s_{\l^-},-s_{\l^+}) \right]}{
    \left[ \Gamma(s_{\l^-},s_{\l^+}) + \Gamma(\mp s_{\l^-},\pm s_{\l^+}) \right] + 
    \left[ \Gamma(\pm s_{\l^-},\mp s_{\l^+}) + \Gamma(-s_{\l^-},-s_{\l^+}) \right]  } \; ,
\end{equation}
and it becomes
\bea
\alp(\bqll)  &=& \frac{ 2
              \sqrt{ 1- \frac{4 m_\l^2 }{m_{B_q}^2} } 
              Re \left[  \frac{m_{B_q}^2}{m_b + m_q} \fps \left( 2 m_\l
                          \faa  - \frac{m_{B_q}^2}{m_b + m_q}~\fpp \right)
                \right]  }
          { 
          \left| 2 m_\l \faa  - \frac{m_B^2}{m_b + m_q}  \fpp \right|^2 
      + (1-\frac{4m_\l^2}{m_{B_q}^2})
            \left|  \frac{m_{B_q}^2}{m_b + m_q} \fps \right|^2 },
  \label{eqn:alp}
\eea
with  $\alp^+ = \alp^- \equiv \alp$.
It is clear that within the SM $\alp(\bqll) \simeq 0$, and
becomes non-zero if and only if $\fps \neq 0$. 
Therefore, this observable would be the best probe to search for
new physics induced by the pseudoscalar type interactions. We also remark that
the dependence on the flavor of the valence quark in $\alp(\bqll)$ 
is tiny, therefore the lepton longitudinal polarization asymmetry is almost
the same for $q = d$ or $q = s$.

\begin{figure}[t]
  \centering
    \begin{minipage}[c]{0.4\textwidth}
     \centering \includegraphics[scale=0.27]{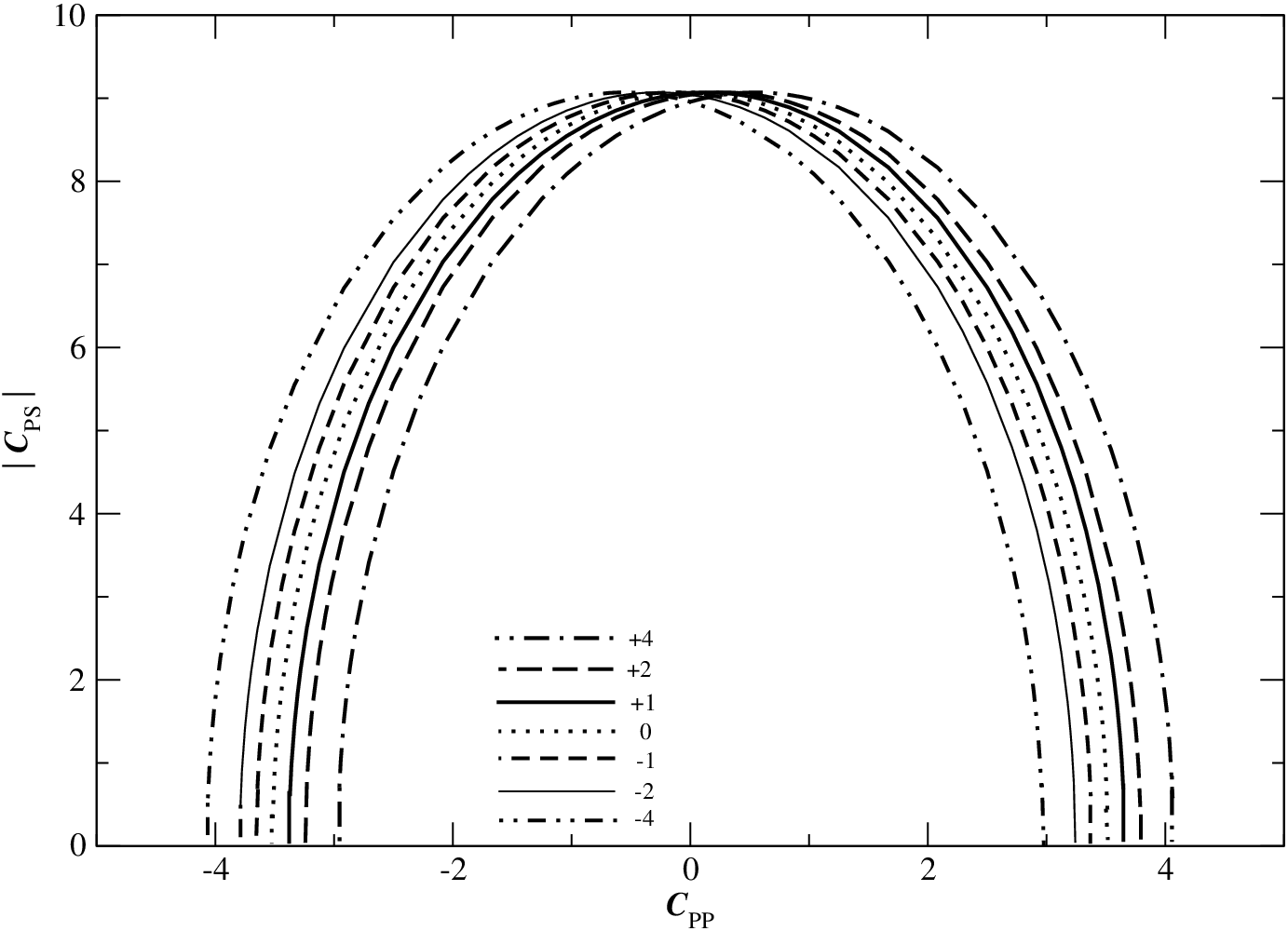}
    \end{minipage}
    \hspace*{5mm}
    \begin{minipage}[c]{0.4\textwidth}
     \centering \includegraphics[scale=0.27]{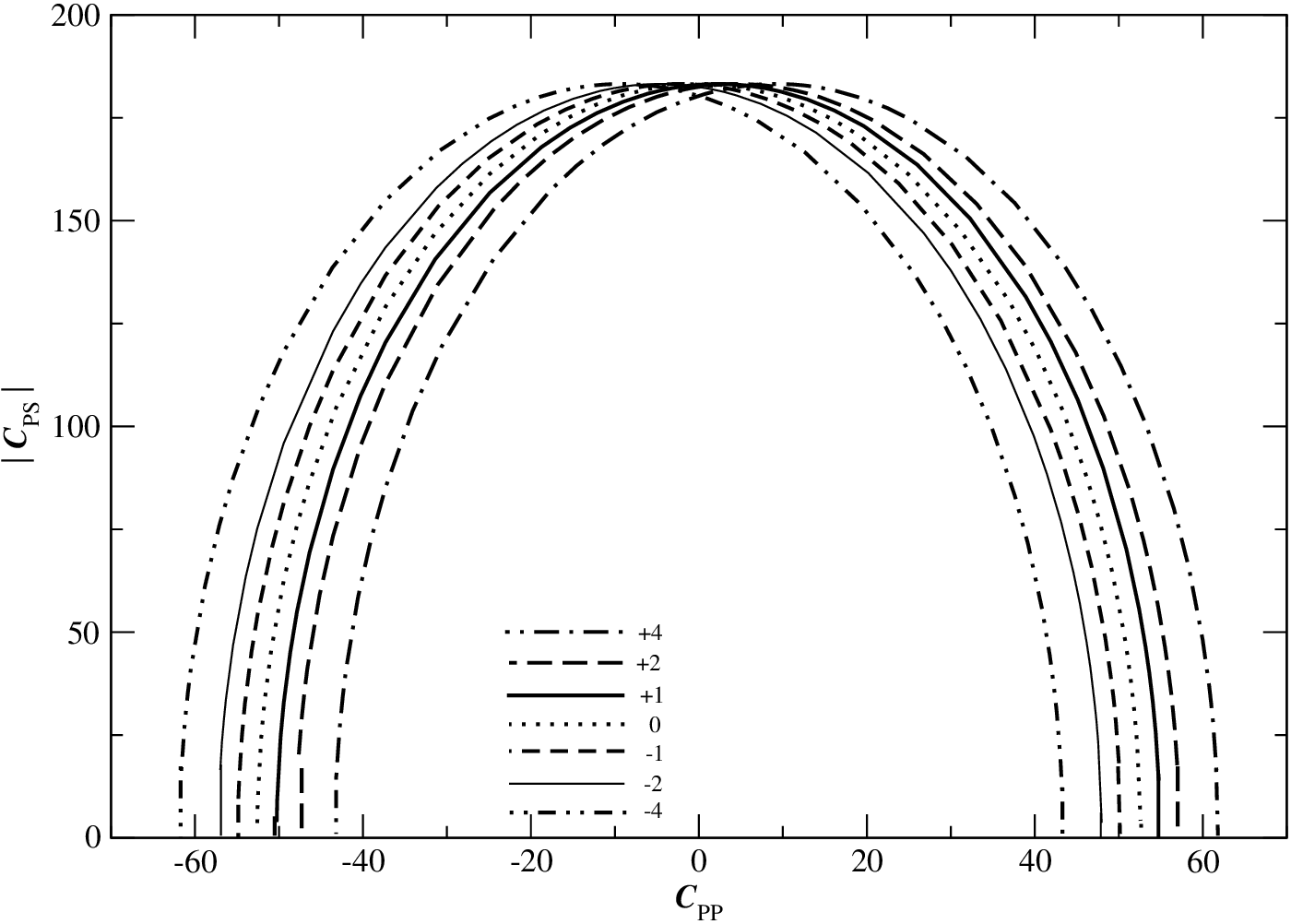}
    \end{minipage}
    \caption{The upper bounds for $\fpp$ vs $|\fps|$
      for $\faa = (-4, -2, -1, 0, +1, +2, +4) \times \faa^{\rm SM}$
      using the experimental bound on $\b(\bsmm)$ (left); and the indirect 
      experimental bound on $\b(\bstt)$ (right).}
    \label{fig:br}
\end{figure}

Before considering physics beyond the SM,
let us briefly review the SM predictions for the processes.
For consistency, the top mass
is rescaled from its pole mass, $m_t = 175 \pm 5$ GeV,
to the $\overline{\rm MS}-$mass, $m_t(\overline{\rm MS}) = 167 \pm 5$ GeV.
For numerical calculations throughout the paper, we use 
the world--averaged values for all other parameters \cite{pdg}, 
{\it i.e.} :
\begin{quote}
$m_{B_q^0} = 5279.2 \pm 1.8$ MeV,
$m_W = 80.41 \pm 0.10$ GeV, $\tau_{B_q^0} = 1.56 \pm 0.04$ (ps)$^{-1}$, 
$m_e = 0.5$ MeV, $m_\mu = 105.7$ MeV, $m_\tau = 1777$ MeV, 
$\sin^2 \theta_W (\overline{\rm MS}) = 0.231$, $\alpha = 1/{129}$, 
$f_{B_d} = 210 \pm 30$ MeV and $f_{B_s} = 245 \pm 30$ MeV \cite{laqcd2}.
\end{quote}
Within the SM and by using the experimental bounds on
the Wolfenstein parametrization
$(A,\lambda) = (0.819\pm0.035,0.2196\pm0.0023)$ 
together with the unitarity of CKM matrix \cite{pdg,ratiobqbq}, we get
\begin{equation}
  \begin{array}{rcl}
  |V_{ts}| & \approx & A \, \lambda^2 = 0.0395 \pm 0.0019 \, , \\
  |V_{td}| & \approx & A \, \lambda^3 
    \sqrt{(1-\rho)^2 + \eta^2} = 0.004 \sim 0.013 \, . 
  \end{array}
\end{equation}
Adopting the next-to-leading order result for $Y(x_{t_W})$  \cite{buras},
and using the central values for all input parameters,
lead to the  following SM predictions,
\begin{eqnarray}
  \b(\bdll) & = & \left\{ 
    \begin{array}{lcl}
  3.4 \times {10}^{-15} \left( \frac{f_{B_d}}{210~MeV}\right)^2 & , & \l = e \\
  1.5 \times {10}^{-10} \left( \frac{f_{B_d}}{210~MeV}\right)^2 & , & \l = \mu \\
  3.2 \times {10}^{-8} \left( \frac{f_{B_d}}{210~MeV}\right)^2  & , & \l = \tau
    \end{array} 
    \right. \, , 
    \label{eqn:tpsm}\\
  \b(\bsll) & = & \left\{ 
    \begin{array}{lcl}
  8.9 \times {10}^{-14} \left( \frac{f_{B_s}}{245~MeV}\right)^2 & , & \l = e \\
  4.0 \times {10}^{-9} \left( \frac{f_{B_s}}{245~MeV}\right)^2 & , & \l = \mu \\
  8.3 \times {10}^{-7} \left( \frac{f_{B_s}}{245~MeV}\right)^2 & , & \l = \tau
    \end{array} 
    \right. \, .
\end{eqnarray}
These predictions should be confronted with the present experimentally
known bounds of $\b(\bqll)$ at $95\%$ CL \cite{cdf},
\begin{eqnarray}
  \b(\bdmm) & < & 8.6 \times {10}^{-7} \, , \\
  \b(\bsmm) & < & 2.6 \times {10}^{-6} \, .
  \label{eqn:bsmmexp}
\end{eqnarray}

\begin{figure}[t]
  \centering
    \begin{minipage}[c]{0.4\textwidth}
     \centering \includegraphics[scale=0.27]{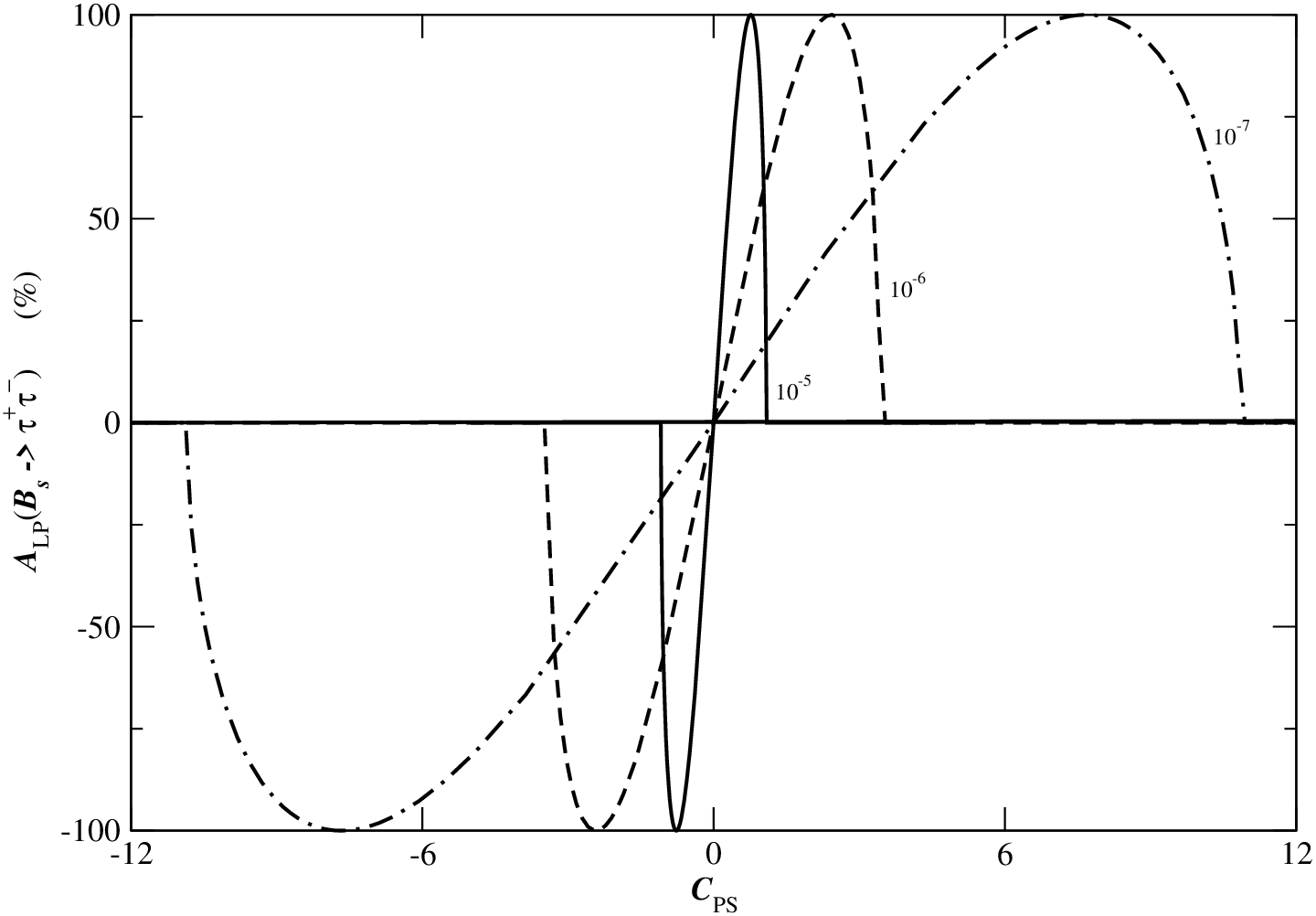}
    \end{minipage}
    \hspace*{5mm}
    \begin{minipage}[c]{0.4\textwidth}
     \centering \includegraphics[scale=0.27]{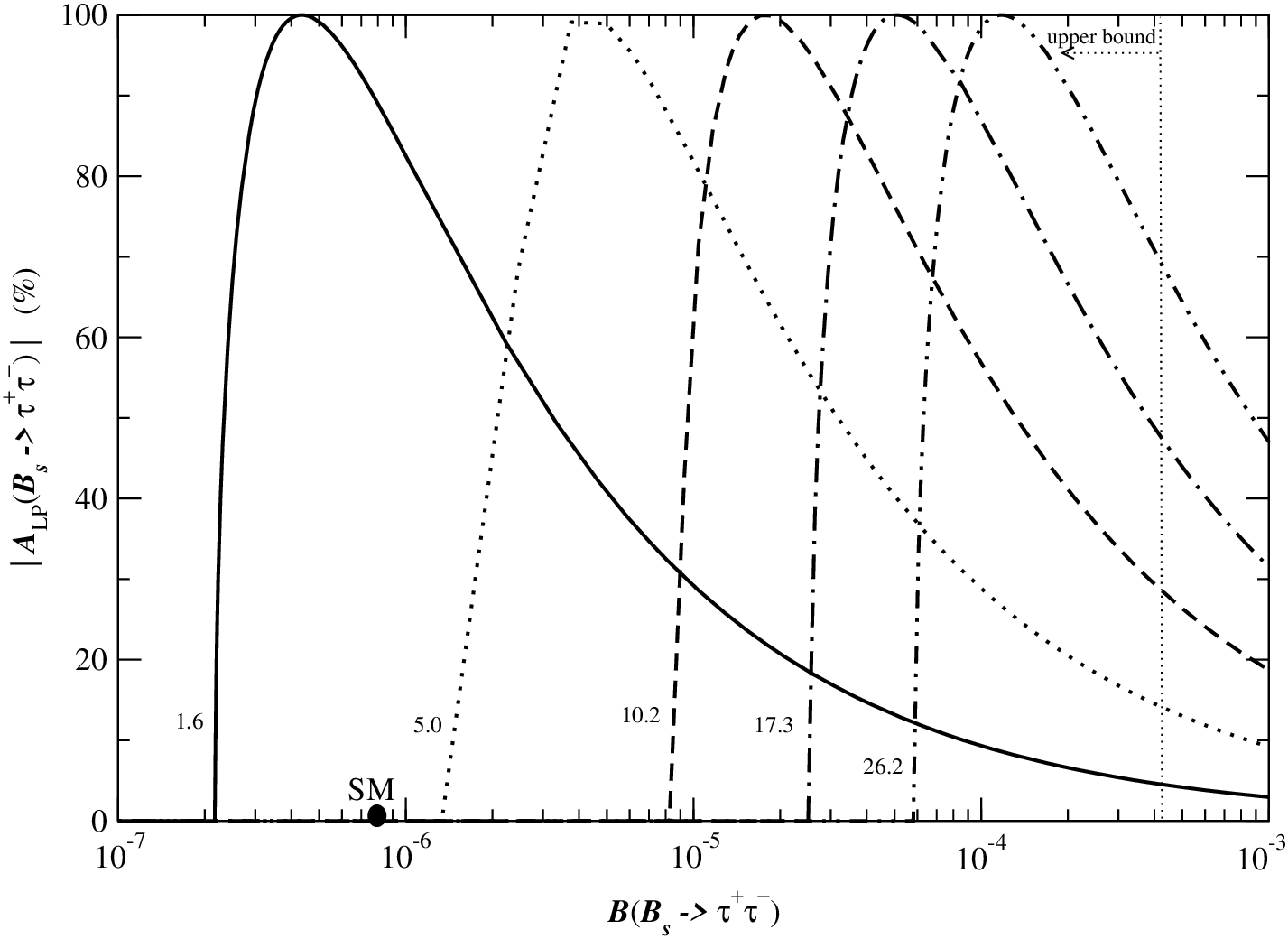}
    \end{minipage}
    \caption{The correlation between $\alp(\bstt)$ and $\fps$
      for various $\b(\bstt) = 10^{-5},10^{-6},10^{-7}$ (left);
      and the correlation between $\alp(\bstt)$
      and $\b(\bstt)$ for various $\fps = 1.6, 5.0, 10.2, 17.3, 26.2$ (right).}
    \label{fig:alpbr}
\end{figure}

To analyze the decay processes and simulataneously find the possible new physics signal,
we first employ  the experimental bound of the 
branching ratio which constraints the coefficients ($\f$'s) 
more strictly after comparing the theoretical predictions with the known 
experimental bounds, $i.e.$
$\b(\bsmm)$ (see Eqs. (\ref{eqn:tpsm})$\sim$(\ref{eqn:bsmmexp})), 
and obtain the allowed region on the $\fps-\fpp$ parameter
space for various values of $\faa$. This is shown in the left-hand-side figure  
of Fig. \ref{fig:br}. In the right-hand-side figure the 
bound is obtained by using the indirect experimental bound 
$\b(\bstt) < 4.3 \times 10^{-4}$ \cite{isidori}. 
Furthermore, suppose that the branching ratio is measured first, then it
must be worth to show a general correlation between the branching ratio
and the longitudinal polarization asymmetry represented by the following
equation, 
\begin{eqnarray}
  \alp(\bqll) &  = & \pm \frac{ 2 a_q \sqrt{ 1 - \frac{4 m_\l^2 }{m_{B_q}^2} }}{\b(\bqll)} \, 
  Re \left[ \frac{m_{B_q}^2}{m_b + m_q} \fps
  \right. \nonumber \\
  & & \left. \times
  \sqrt{\frac{\b(\bqll)}{a_q} - \left( 1 - \frac{4 m_\l^2 }{m_{B_q}^2} \right)
    \left| \frac{m_{B_q}^2}{m_b + m_q} \fps \right|^2 }
  \right] \, ,
\end{eqnarray}
by eliminating $\faa$ and $\fpp$ in Eqs. (\ref{eqn:br}) and (\ref{eqn:alp}),
where the constant $a_q$ is defined as
\begin{equation}
a_q \equiv \frac{G_F^2 \alpha^2}{64 \pi^3} \,  
  \left| V_{tq}^\ast V_{tb}  \, \right|^2 \tau_{B_q} f_{B_q}^2 m_{B_q} \, 
 \sqrt{ 1 - \frac{4 m_\l^2 }{m_{B_q}^2} } \, .
\end{equation}
This  is depicted in Fig. \ref{fig:alpbr}. The left-hand-side figure
shows a correlation between $\alp(\bstt)$ and $\fps$ for 
various $\b(\bstt)$, while 
the right-hand-side one is between $\alp(\bstt)$ and $\b(\bstt)$
for various $\fps$. 

\begin{figure}[h]
    \centering
    \includegraphics[scale=0.5]{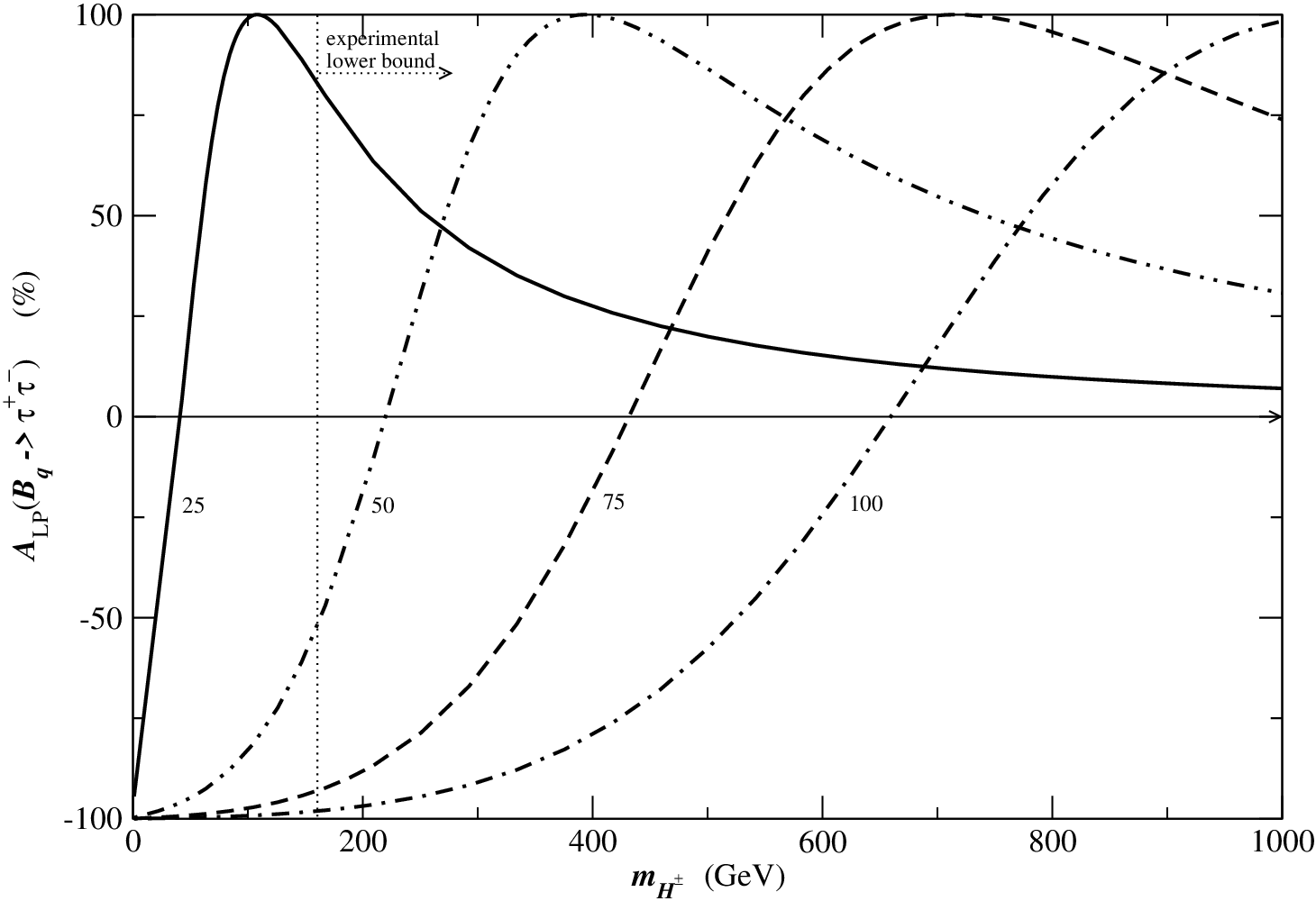}
    \caption{The longitudinal polarization asymmetry of $\tau$'s,
      $\alp(\bqtt)$, as a function of $m_{H^\pm}$
      for various $\tan \beta = 25, 50, 75, 100$. }
    \label{fig:alpbrhdm}
\end{figure}

As a specific example for the case in which $\fps$ is non-zero, we adopt
the type II 2-Higgs-doublet models (2HDM-II).
In this model
$$\faa^{\rm 2HDM-II} = \faa^{\rm SM} ,$$
while\footnote{We take the latest results calculated in \cite{logan2} 
     by neglecting the subleading terms 
    in $\tan \beta$. Note that the results are consistent with 
    \cite{kruger} if one drops the contributions from trilinear coupling. }
\begin{equation}
  \fps^{\rm 2HDM-II} = \fpp^{\rm 2HDM-II} = 
  \frac{m_\l (m_b + m_q)}{4 M_W^2 \sin^2 \theta_W} 
  \, \tan^2 \beta \frac{\ln x_{H^\pm t}}{x_{H^\pm t} - 1} \; ,
\end{equation} 
at large $\tan \beta$ limit \cite{logan2,huang2hd,kruger}, and 
$x_{H^\pm t} = ({m_{H^\pm}}/{m_t})^2$. 
Some particular cases in the right-hand-side figure of
Fig. \ref{fig:alpbr} can be realized by, for instance,
\begin{quote}
$(m_{H^\pm},\tan \beta) = (200 \, {\rm GeV},40)$ for $\fps = 1.6$,
$(200 \, {\rm GeV},70)$ for $\fps = 5.0$, 
$(200 \, {\rm GeV},100)$ for $\fps = 10.2$, 
$(200 \, {\rm GeV},130)$ for $\fps = 17.3$, 
$(200 \, {\rm GeV},160)$ for $\fps = 26.2$.
\end{quote}

Finally, in Fig. \ref{fig:alpbrhdm} we show the
dependences of $\alp(\bqtt)$ on $m_{H^\pm}$ and $\tan \beta$.
For the real experimental analyses, 
we recommend $\bstt$ decays because the energy of final $\tau$'s 
is high enough to decay further to energetic secondary particles, so 
their longitudinal polarization  may be well measured in hadronic $B-$factories. 
Although the $\tau$'s are difficult to be reconstructed in 
hadronic background, we need precisely 
such reconstruction from their decay products that 
allows  measurements of the longitudinal polarization of $\tau$'s. 

In conclusion we have considered a general analysis exploring
the  longitudinal polarization asymmetry of leptons in the $\bqll$ decays. 
We have
shown that this observable would provide a direct measurement of the 
physics of scalar and pseudoscalar type interactions. 
We also note that more information
about these new interactions can be obtained by combining the present
analysis with the other observables from $B\rightarrow X_q \l^+ \l^-$ \cite{bsll}. \\

\noindent
We thank G. Cvetic  and D. London for careful reading of the manuscript and their
valuable comments.
The work of C.S.K. was supported
by Grant No. 2001-042-D00022 of the KRF.
The work of T.Y. was supported in part by the US Department of Energy
under Grant No.DE-FG02-97ER-41036. 

\newpage


\end{document}